\documentclass[pre]{revtex4-2}

\usepackage[T1]{fontenc}
\usepackage{amsfonts}
\usepackage{amsmath}
\usepackage{amssymb}
\usepackage[dvips,pdftex]{graphicx}
\usepackage{hyperref}
\usepackage{color}
\usepackage{float}
\usepackage{enumitem}

\begin{document}

\title{Uncertainties in Signal Recovery from Heterogeneous and Convoluted Time Series\\ with Principal Component Analysis}
\author{Mariia Legenkaia$^{1,2}$, Laurent Bourdieu$^{1,*}$, R\'emi Monasson$^{2,*}$}
\address{$^1$ Institut de Biologie de l’Ecole Normale
Supérieure (IBENS), Ecole Normale Supérieure, CNRS,
INSERM, Université PSL, Paris F-75005, France \\$^2$ Laboratoire de Physique de l'ENS, PSL \& CNRS-UMR8023, Sorbonne Universit\'e, 24 Rue Lhomond, 75005 Paris, France\\
$*$ Equal contributions}

\begin{abstract}
Principal Component Analysis (PCA) is one of the most used tools for extracting low-dimensional representations of data, in particular for time series. Performances are known to strongly depend on the quality (amount of noise) and the quantity of data. We here investigate the impact of heterogeneities, often present in real data, on the reconstruction of low-dimensional trajectories and of their associated modes. We focus in particular on the effects of sample--to--sample fluctuations and of component-dependent temporal convolution and noise in the measurements. We derive analytical predictions for the error on the reconstructed trajectory and the confusion between the modes using the replica method in a high-dimensional setting, in which the number and the dimension of the data are comparable. 
We find in particular that sample-to-sample variability, is deleterious for the reconstruction of the signal trajectory, but beneficial for the inference of the modes, and that the fluctuations in the temporal convolution kernels prevent perfect recovery of the latent modes even for very weak measurement noise. Our predictions are corroborated by simulations with synthetic data for a variety of control parameters.
\end{abstract}

\maketitle

\section{Introduction} 

Principal component analysis (PCA) is, without any doubt, one of the most used statistical tools for dimensionality reduction \cite{Joliffe2002,Jolliffe2016,Greenacre2022}. Applications abound in science and beyond, for instance, for multidimensional scaling, {\em i.e.} the visualization of complex data in (generally) two or three dimensions. PCA has been used in a wide variety of research fields from image classification and compression \cite{Rodarmel2002,Du2007} and face recognition \cite{Paul2012} to quantitative finance \cite{Ghorbani2020}, neuroscience \cite{Cunningham2014,Gallego2017,Williamson2019}, genetics and genomics \cite{Abraham2014,Alter2000, Tsuyuzaki2020}, and atmospheric science \cite{Obukhov1947,Lorenz1956,Preisendorfer1988}. It has also been extended to the social sciences, where it is known as correspondence analysis \cite{Benzecri1973}.

Informally speaking, PCA looks for a few groups of variables that tend to vary in a correlated way in the data. The projections of the high-dimensional data points on these groups, called modes, define a low-dimensional representation of the data. This approach is particularly useful for time series associated to the dynamics of complex systems, as it obtains low-dimensional trajectories of their most salient features. PCA is therefore used in many domains, such as neuroscience, where applications to neural activity recordings provide effective trajectories representing the accomplishment of tasks by animals \cite{Cunningham2014}. 

As for any statistical method, it is important to assess the accuracy of PCA in simple settings, where precise analysis can be carried out.  In this context, much attention has been brought in the physics \cite{Watkin1994,Reimann1996,hoyle2004} and in the mathematics \cite{Johnstone2001, Baik2004} communities to simple data models, the covariance of which includes few symmetry-breaking directions. One illustration is the spiked covariance model, in which the structure of the data is isotropic along all components in a high-dimensional space, except along one direction. The issue is then to know whether PCA is able to identify this special direction, depending on the quantity and quality of the data. More precisely, in the spiked covariance model, one considers $T$  data points ${\bf s}_t$, $t=1,...,T$, in dimension $N$, whose components are given by
\begin{equation}\label{spike}
    s_{i,t} =  x_t \; e_i + z_{i,t}\ ,
\end{equation}
where ${\bf e}=(e_1,e_2,...,e_N)$ is the vector defining the special direction, $x_t$ is the coordinate at time $t$ along this direction, and $z$ is a white noise process, with uniform variance $\sigma^2$ for all components $i$. An important result is that, in the high-dimensional setting where both $T$ and $N$ are sent to infinity at fixed ratio $r=N/T$, the top mode identified by PCA, ${\bf v}$, is orthogonal to the special direction $\bf e$ if $\text{var}(x)/\sigma^2 < r$. In other words, if the time excursion of the coordinate compared to the noise in the data is not large enough, then the special direction cannot be recovered. Above this critical threshold, recovery of the true direction is possible and improves as more and more data become available. Projecting the data point ${\bf s}_t$ along $\bf v$ then allows for estimating the latent coordinate $x_t$. This result holds in the case of $K>1$ modes ${\bf e}^{(1)}, {\bf e}^{(2)}, ..., {\bf e}^{(K)}$ to be extracted from the data, with a sequence of critical thresholds defined by the variances of the coordinates along them, as long as $K$ is finite compared to $N$ and $T$.

While this result is of theoretical interest and can be used to estimate when sampling is sufficient to extract signal in some simple contexts, it is of limited applicability due to the oversimplified nature of the data model in Eq.~\eqref{spike}. A fundamental feature of real data is the presence of non-homogeneities, either intrinsic to the system or resulting from the measurement process. More precisely, we want to acknowledge that

\begin{enumerate}[{label=(\arabic{*})}]
    \item The measurement noise is generally not uniform across components. This is expected to be the case for high-dimensional measurements, in which the different components account for various characteristics of the system. From a mathematical point of view, the variance of $z_{i,t}$ in Eq.~\eqref{spike} may thus strongly vary with $i$. 
    \item Physical measurement devices, which we should hereafter refer to as probes, are not instantaneous. In other words, one does generally not have access to the fully resolved in time state of the system, but to some time-convoluted version. Probes associated to different features $i$ of the system may behave in different ways, so the convolution kernel is expected to vary with the component $i$.
    \item In many applications of interest, data are collected from different sources or from different realizations of the same experiment \cite{Oba2007,Shi2024}. However, due to the specificity of the sources or to the intrinsic variability or to a lack of control of the initial or external conditions, the system coordinate $x_t$ may substantially fluctuate from one sample to another. 
\end{enumerate}
In this work, we introduce a new data model, which extends over the standard spiked covariance model and includes the three sources of complexity above. Yet, our model is amenable to exact analysis. We derive, using the replica method of the statistical mechanics of disordered systems, exact expressions for observables of interest in an asymptotic regime. We study in particular the accuracy of the reconstructed trajectory $x_t^{(1)}, x_t^{(2)},...,x_t^{(K)}$ and of the inferred modes ${\bf e}^{(1)}, {\bf e}^{(2)}, ..., {\bf e}^{(K)}$. The latter quantity is important to understand how the signal is encoded at the microscopic level, as some component $i$ may strongly participate in some modes and much less in other ones.

Our paper is organized as follows. In Section II, we introduce the different ingredients of the data model and the definitions of the observables. In Section III, we present our statistical mechanics calculation and the main analytical results. In Section IV, we compare our theoretical predictions to numerical data (application of PCA to synthetic data). Last of all, some conclusions and perspectives are presented in Section V.

\section{Data model and errors} 
\label{sec:model}

The data model considered in this work is an extension of the spiked covariance model, see Eq.~\eqref{spike}, in which realistic aspects of the intrinsic and measurement variability are taken into account. We introduce the model and its ingredients in Section \ref{sec:modeldef}, and define the observables of interest in Section \ref{sec:observable}.

In the following, a data point is a $N$-dimensional real-valued vector ${\bf s}_t$, where $t=1,...,T$ is the time index. The components of the data points are labeled by $i=1,...,N$. We refer to Table~\ref{tab:data} for these notations and the following ones.

\subsection{Model definition}
\label{sec:modeldef}

\subsubsection{Latent modes and (mean) latent signal}
\label{sec:slowlatent}

The core assumption in our data model is that, despite being of very high dimension $N$, only few directions, $K$, carry out informative signals. We refer to these $K$ orthogonal directions as the $N$-dimensional vectors $\mathbf{e}^{(k)}$, with $k=1,...,K$. We assume that the vectors have squared norms equal to $N$, so that their components are typically of the order of unity. Notice that components $i$ may participate in some modes, {\em e.g.} $|e_i^{(1)}| \sim 1$, and be irrelevant for other ones, {\em e.g.} $|e_i^{(k\ge 2)}| \simeq 0$. 

In the $K$-dimensional latent space, the system dynamics is described by the signal coordinates $x_t^{(k)}$. We assume that
\begin{itemize}
    \item the signal coordinates are centered, {\em i.e.} the average over time of $x_t^{(k)}$ vanishes;
    \item the signal trajectory is smooth over time, {\em i.e.} the signal significantly varies over a correlation time $\tau_{signal}$, with $1 \ll \tau_{signal} < T$. 
\end{itemize}

\begin{figure}
    \centering
\includegraphics[width=1\linewidth]{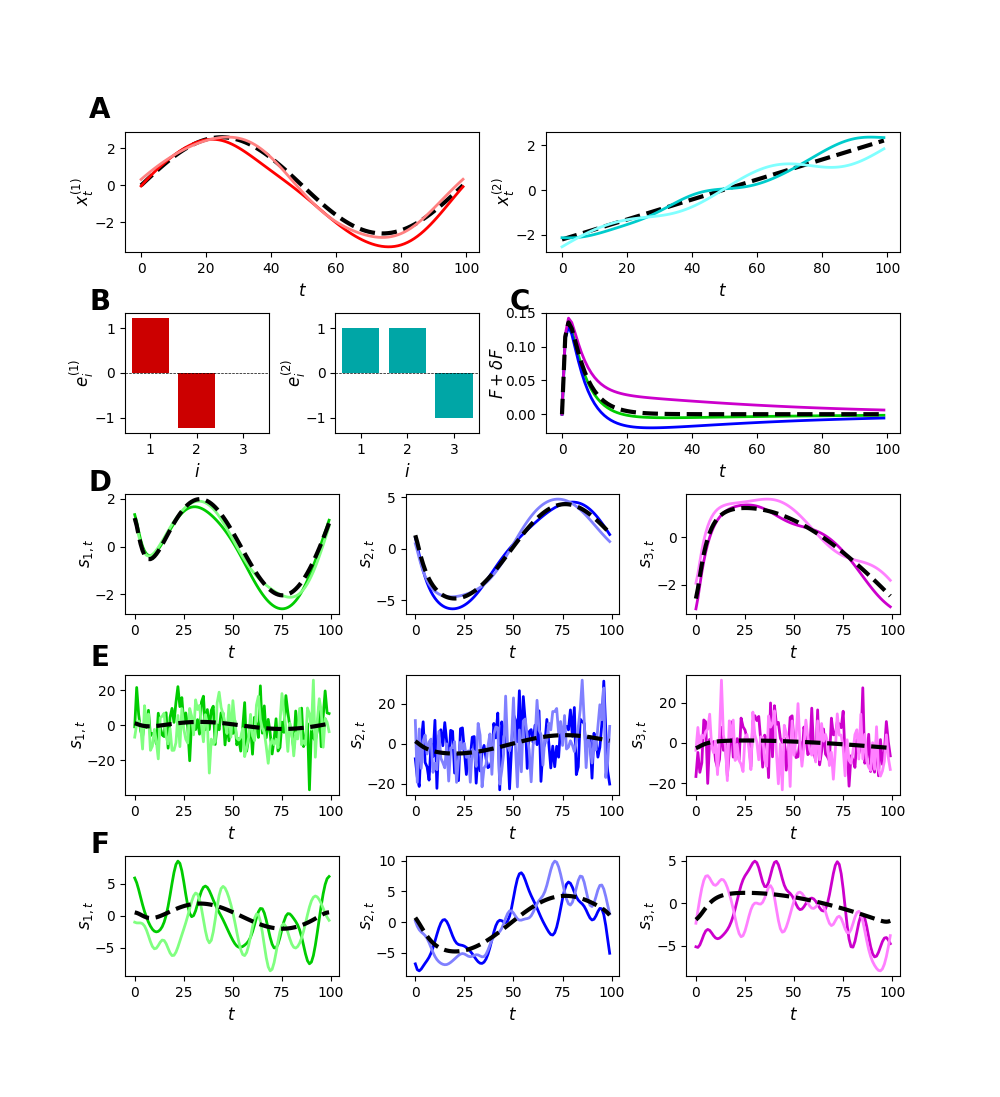}
\caption{{\bf Ingredients of the data model} illustrated in the case of $K=2$ modes. We show 2 data samples for $N=3$ components and $T=100$ times steps.
{\bf A.} The signal coordinates $x_t^{(k)}$ for modes $k=1$ (left, red) and $k=2$ (right, cyan) are displayed with black dashed lines. For each one of the two samples, we show $x_t^{(k)}+ \delta x_t^{(k)}$; $\delta x$ is generated as a Gaussian noise with variance $\xi^2=0.01$ and temporal correlation $\Delta$ decaying exponentially over $\tau_{\delta x} = 10$.
{\bf B.} Values of $e_i^{(k)}$ for the three components $i=1,2,3$ (left \& red, $k=1$; right \& cyan, $k=2$). 
{\bf C.} Measurement convolution kernels. Dashed black lines: $F_\tau$. Colored lines: $F_\tau+\delta F_{i,\tau}$ for $i=1$ (green), 2 (blue), and 3 (purple).
{\bf D.} Data components $s_{i,t}$ vs. time step $t$ for $i=1$ (left), 2 (middle), and 3 (right) after convolution of with the measurement convolution kernels.
{\bf E.} Same as previous row, with additive fast noise $z_{i,t}$.
{\bf F.} Same as previous row, after smoothing with a Gaussian $G$ kernel with decay time $\tau_z = 3$. Note the different vertical scales in panels {\bf D}, {\bf E}, {\bf G}. 
}
\label{fig:modeldata}
\end{figure}

We illustrate the notions of latent modes and latent signal in Figure~\ref{fig:modeldata}.

\subsubsection{Sample-dependent variations of latent signal}

Repeated measurements of the system dynamics give information about the intrinsic variability of its low-dimensional trajectory in the latent space. Besides the fast noise due to measurements, we assume that their exist a slow, sample-dependent noise, which we model as a Gaussian process around the mean signal $x$ defined above. More precisely, we assume that, for each sample, the latent signal is actually $x_t^{(k)} + \delta x_t^{(k)}$, where 
\begin{itemize}
    \item the fluctuations of signal coordinates are centered, {\em i.e.} the average over samples of $\delta x_t^{(k)}$ vanishes for all $t$ and $k$;
    \item the fluctuations are drawn according to a multivariate Gaussian distribution, with covariance matrix
    \begin{equation}
        \text{Cov} \left( \delta x^{(k)}_t, \delta x^{(k')}_{t'} \right) = (\xi^{(k)})^2 \; \delta_{k,k'} \;\Delta_{t,t'} \ ,
    \end{equation}
    where $\xi ^{(k)}$ is the standard deviation along mode $k$. We expect $\Delta$ to decay over a time scale comparable to $\tau_{signal}$. 
\end{itemize}
Examples of fluctuations in the latent signal are shown in Figure~\ref{fig:modeldata}. 

\subsubsection{Component-specific convolution of the signal in the measurement process}

The signal components are generally not directly accessible for measurements. We hereafter assume that they can be read out by $N$ linear probes, one for each component. However, these probes cannot adapt to the instantaneous value of the signal, and report its convolution with a time- and component-dependent kernel, $F_{i,\tau}$, where $\tau\ge 0$ represents the delay. Ideal probes, instantaneously reporting the signal, would correspond to $F_{i,\tau}=\delta_{\tau,0}$. Examples of convolution kernels are shown in Figure~\ref{fig:modeldata}.

We assume that
\begin{itemize}
    \item the average over all components $i$ of the probe convolution kernels is $F_\tau$. This mean kernel smoothly decays with $\tau$ over the typical integration time of the probes;
    \item the fluctuations of the convolution kernel from probe to probe are drawn according to a multivariate Gaussian distribution, with covariance matrix
    \begin{equation}\label{eq:defxi}
        \text{Cov} \left( \delta F_{i,\tau},  \delta F_{i',\tau'} \right) = \delta_{i,i'} \, \Xi_{\tau,\tau'}   \ .
    \end{equation}
    Notice that $\Xi$ is not a simple function of the difference $\tau'-\tau$, as it depends on the overall shape of the convolution kernel distributions. In the following, we also need to consider the tensor
     \begin{equation}
        \mathcal{X}^{(k_1,k_2)}_{t_1,t_2} = \sum_{\tau_1,\tau_2} x_{t_1-\tau_1}^{(k_1)} \, \Xi_{\tau_1,\tau_2}\,  x_{t_2-\tau_2}^{(k_2)}  \ .
    \end{equation}
\end{itemize}

\subsubsection{Fast measurement noise}

In addition, the measurement is plagued by additive noise $z$, which we assume to be uncorrelated in time and across the components. The noise is on average equal to zero, and the variance may depend on the component $i$. Hence, the covariance of this fast noise is given by
    \begin{equation}
        \text{Cov} \left( z_{i,t}, z_{i',t'}  \right) = \sigma_i^2 \; \delta_{i,i'} \;\delta_{t,t'} \ ,
    \end{equation}
where $\sigma_i$ is the standard deviation along component $i$. We provide in Figure~\ref{fig:modeldata} an illustration of the effects of this fast measurement noise. 

\subsubsection{Raw data and smoothing process}

Gathering all the contributions above, we obtain the following expression for the raw data components:
\begin{eqnarray}\label{eq:dataraw}
    s_{i,t} =  \sum\limits_{\tau}  (F_{\tau} + \delta F_{i,
    \tau}) \sum\limits_k\left( x^{(k)}_{t-\tau} + \delta x^{(k)}_{t-\tau} \right) e_i^{(k)} + z_{i,t} \ .
\end{eqnarray}
Notice that the variances $\big(\xi^{(k)}\big)^2$ and $\sigma_i^2$ above describe the distributions of, respectively, $\delta x$ and $z$ for a single sample. When averaging data over multiple, say $S$, samples, $\Delta$ and $\sigma_i^2$ should be rescaled by $1/S$.

While raw data are accessible and could be processed by PCA, we can take advantage of the `slow' latent dynamics, see Section~\ref{sec:slowlatent}, to filter out the fast measurement noise. In practice, we introduce a temporal smoothing kernel $G$, convoluting the data over a time scale $\tau_z$ smaller than $\tau_{signal}$, and define the smoothed data
\begin{eqnarray}\label{eq:datasmoothed}
    \bar s_{i,t} = \sum\limits_{\tau'} G_{\tau'} \; s_{i,t-\tau'} \ ,
\end{eqnarray}
where $s$ was defined in Eq.~\eqref{eq:dataraw}. For the sake of simplicity, we impose periodic boundary conditions for the convolution in the equation above, {\em i.e.} $t-\tau'$ is computed modulo $T$ in Eq.~\eqref{eq:datasmoothed}. Notice that, with this smoothing process, the covariance of the smoothed noise $\bar z_{i,t} = \sum _\tau G_\tau z_{i,t-\tau} $ becomes
\begin{equation} \label{eq:z}
     \text{Cov} \left( \bar z_{i,t}, \bar z_{i',t'}  \right) = \bar \sigma_i^2 \; \delta_{i,i'} \;Z_{t,t'} \quad \text{with} \quad \bar \sigma_i^2 = \sigma_i^2\, \sum _{\tau} G_\tau^2 \quad \text{and}\quad Z_{t,t'}=\frac{\sum_{\tau } G_{t-\tau} \, G_{t'-\tau} }{\sum _{\tau} G_\tau^2 }\ .
\end{equation}
We hereafter use the $\bar{\cdot}$ notation to indicate convolution with the smoothing kernel. This smoothing can be applied to vectors or to matrices. As an illustration,
\begin{equation} \label{eq:smooth}
    \bar x_{t}^{(k)}  = \sum_{\tau} G_{\tau} \;  x_{t-\tau}^{(k)}  \quad \text{and} \quad
   \bar\Delta_{t,t'} = \sum_{\tau,\tau'} G_{\tau} \, \Delta_{t-\tau,t'-\tau'}\,  G_{\tau'} 
\end{equation}

\begin{table}
\begin{tabular}{|c|c|c|}
    \hline
    Parameter & Meaning & Further description  \\
    \hline
    $N$ & Number of components & \\ 
    $T$ & Number of time steps & \\
    $K$ & Dimensionality of latent activity & \\
    $i$ & Component index & in range $1, \dots, N$ \\
    $t$ & Time index & in range $1, \dots, T$ \\
    $k$ & Index of a latent mode & in range $1, \dots, K$ \\
    $e_i^{(k)}$ & $k$-th latent mode  & Normalized: $\sum_i (e_i^{(k)})^2 = N$ \\
    $x_t^{(k)}$ & Time-dependent signal along $k$-th mode & Sample invariant \\
    $\delta x_t^{(k)}$ & Sample-dependent fluctuations of signal  & $\text{Cov} \left( \delta x^{(k)}_t, \delta x^{(k')}_{t'} \right) = (\xi^{(k)})^2 \delta_{k,k'} \Delta_{t,t'}$ \\
    ${F}_\tau$ & Mean measurement convolution kernel of signal & \\
    $\delta F_{i,\tau}$ & Additive contribution to measurement & $\text{Cov} \left( \delta F_{i,\tau},  \delta F_{i',\tau'} \right) = \delta_{i,i'} \, \Xi_{\tau,\tau'}$  \\ &  convolution kernel for unit $i$ &  \\
     $z_{i,t}$ & Measurement noise for component $i$ & $\text{Cov} \left( z_{i,t}, z_{i',t'} \right) = \sigma_i^2 \, \delta_{i,i'} \, \delta_{t,t'}$ \\
       $G_{\tau}$ & Temporal smoothing kernel & Gaussian with width $\tau_z$ \\
       $\bar z_{i,t}$ & Smoothed measurement noise for component $i$ & $\text{Cov} \left( \bar z_{i,t}, \bar z_{i',t'} \right) =\bar \sigma_i^2 \, \delta_{i,i'} \, Z_{t,t'}$ \\
    \hline
\end{tabular}
\caption{Notations in the data model. Left column: variables appearing in the raw and smoothed data, see Eqs.~\eqref{eq:dataraw},\eqref{eq:datasmoothed}. Middle: definition of the variables. Right: Additional information, see text. }
\label{tab:data}
\end{table}

\subsection{Observables of interest}
\label{sec:observable}

After applying the smoothing process to the raw data, see Eq.~\eqref{eq:datasmoothed}, we compute the covariance matrix
\begin{equation}\label{def:covmat}
    C_{i,j}= \frac 1T \sum_{t=1}^T \bar s_{i,t}\, \bar s_{j,t} - \frac 1T \sum_{t=1}^T \bar s_{i,t}\times \frac 1T \sum_{t=1}^T \bar s_{j,t} \ .
\end{equation}
If multiple samples are available, $\bar s_{i,t}$ in the equation above is replaced with its average over the samples (at fixed $i$ and $t$).
We then diagonalize $C$ and obtain the $K$ top eigenvalues and associated modes $\mathbf{v}^{(k)}$, with $k=1,\ldots, K$. 

Once the principal modes have been determined, our estimate for the signal coordinates is simply obtained from the  projections of the raw data:
\begin{equation}\label{eq:signalinf}
y_t^{(k)}= \frac {1}{N} \sum _{i=1}^N v_i^{(k)}\,  s_{i,t} \ ,
\end{equation}
where the normalization factor comes from the fact that the $\mathbf{v}^{(k)}$'s are normalized to $\sqrt N$.

\subsubsection{Error of the estimation of the shape of the neural trajectory}

It is important to understand how closely the reconstructed trajectories match the true latent dynamics of the system. In terms of the model notation that we have introduced before, we would like to know how much $y^{(k)}_t$ differs from $x^{(k)}_t$. We therefore introduce the covariance matrix
    \begin{equation}
        \epsilon^{(k,k')} = \frac{1}{T} \sum\limits_{t=1}^T \big( y_t^{(k)} -x_t^{(k)}\big) \big(y_t^{(k')} -x_t^{(k')}\big) \label{epsilon_definition}  .
    \end{equation}
The diagonal elements $\epsilon^{(k,k)}$ correspond to the average squared distance between the ground-truth and inferred signals, see Eq.~\eqref{eq:signalinf}. Strong off-diagonal terms indicates that PCA has mixed up the different signal modes.

\subsubsection{Error on the estimate of the  principal components}

PCA determines the directions $\mathbf{e}^{(k)}$, along which the variability in the data is greatest. The weights (or loadings) $e_i^{(k)}$ indicate how much the component $i$ participates in the mode. If several directions have high loads on the same mode, they tend to be correlated (positively or negatively), contributing to the same underlying collective feature in the data.
    
To measure how well the inferred modes $\mathbf{v}^{(k)}$ obtained with PCA approximate the ground-truth latent modes $\mathbf{e}^{(k)}$, we introduce the $K\times K$--covariance matrix
\begin{eqnarray}\label{eq:defrho}
  \rho^{(k,k')} = \frac{1}{2N} \sum\limits_{i=1}^N  \left( v^{(k)}_i  - e^{(k)}_i \right)\left( v^{(k')}_i  - e^{(k')}_i \right) \ .
\end{eqnarray}
The diagonal term $k=k'$ correspond to the squared norm of the difference vector $\mathbf{d} ^{(k)}= \mathbf{v}^{(k)}-\mathbf{e}^{(k)}$; the off-diagonal terms measure the dot products between the difference vectors.
Notice that the factor $\frac 12$ is introduced for convenience reasons, as it allows us to simplify the final expression, see Section \ref{sec:observ}.

\section{Analytical calculation of the reconstruction errors} 
\label{sec:calcul}

\subsection{Statistical-mechanics formalism}

\subsubsection{Measure over the sets of $K$ vectors} 

Our goal in this section is to derive the average values of $\epsilon$ and $\rho$ as functions of the different parameters defining the data (measurements, processing, ...).  To do so, we will resort to the replica method of the statistical physics of disordered systems, for an introduction see \cite{Steinberg2024}. We introduce an energy function over the $K\times N$-dimensional space vectors $\bf v^{(k)}$, $k=1,...,K$. The energy is defined through the quadratic form
\begin{eqnarray}
E\big[\{\mathbf{v}^{(k)}\}, \{s_{i,t}\}\big] =-\frac{1}{N^2} \sum\limits_{k=1}^K \sum_{i,j=1}^{N} v_i^{(k)} \, C_{ij} \, v_j^{(k)}  \label{eq:energy}
\end{eqnarray}
The energy is minimized by a set of  $K$ orthonormal vectors $v^{(k)}$ that align with the maximal-variance directions, that is, along the first $K$ principal components of the covariance matrix $C$, which is computed from the data, see Eq.~\eqref{def:covmat}. 
We therefore  introduce the following partition function:
\begin{eqnarray}
Z \big(\{s_{i,t}\}\big) &=& \int \prod\limits_k d \mathbf{v}^{(k)}  \prod\limits_{k,t} \delta \left( \sum_i (v_i^{(k)})^2-N\right) \prod_{k<k'} \delta \left( \sum_i v_i^{(k)} v_i^{(k')} \right)  \exp\left( - \beta \,T\,  E\big[\{\mathbf{v}^{(k)}\}, \{s_{i,t}\}\big] \right) \ .
\label{partition_function}
\end{eqnarray}
where $\beta$ plays the role of an inverse temperature. We hereafter denote by $\langle \cdot\rangle$ the average over the Boltzmann measure over the set of $K$ vectors implicitly defined in $Z$ above, and by ${\lceil\cdot\rceil}$ the average over the quenched disorder, {\em i.e.} the data $s_{i,t}$, the sample-to-sample variations $\delta x_t^{(k)}$, and the convolution kernel fluctuations $\delta F_{i,\tau}$. We will eventually send $\beta \to \infty$ to concentrate the measure on the optimal set of vectors. In this zero-temperature limit, the ground-state energy coincides with the sum of the top $K$ eigenvalues of the covariance matrix.
In addition, we may introduce small conjugated forces to the observables $\rho^{(k,k')}$ and $\epsilon^{(k,k')}$ above; differentiating with respect to these forces allows us to compute the average values of the observables. 

\subsubsection{The double infinite-size limit}

In practice, the calculation of the typical value of the partition function can be done with the replica method. The set of $K$ vectors is copied $n$ times, and this replicated partition function is averaged over the data, inducing effective interactions between the replicas. Due to the mean-field nature of the resulting theory (lack of spatial structure and low-rank interactions), it can be solved in the large-size limit through a saddle-point estimation. We expect replica symmetry to hold due to the quadratic nature of the energy $E$. Analytic continuation to $n\to 0$ provides us with the mean value of $\log Z$.

The whole calculation is standard and closely follows the lines of \cite{Reimann1996,hoyle2004}; the main novelty is the presence of the variable-dependent convolution kernels $\delta F$. To be able to carry out the Gaussian integrals over the fluctuations in the measurement convolution kernels and in the latent signal, we need to decouple these terms. In other words, we neglect the multiplicative term $\delta F_{i,\tau}\,\delta x^{(k)}_{t-\tau}$ in Eq. \eqref{eq:dataraw}, which is assumed to be small, and keep only the constant and linear terms in $
\delta F, \delta x$. Detailed calculations can be found in \cite{Legenkaia2024}.

We stress that the saddle-point approach allows us to estimate the ground-state energy and the observables of interest in the double large-size limit $N,T\to \infty$ at fixed ratio $\alpha=T/N$. In this asymptotic setting, the average values of intensive observables, which {\em a priori} depend on both $N$ and $T$, become functions of $\alpha$ only. Some care has then to be brought to the scaling of the model-defining parameters to ensure that this infinite-size limit is non-trivial. As an illustration, consider the additive noise $z_{i,t}$ on each component in Eq. \eqref{eq:z}. If the variance $\sigma^2_i$ of the noise is finite (for all $i$) when $N\to\infty$, then the inference problem becomes trivial: the strength of the signal increases linearly with the number $N$ of components, while the effect of the noise terms, uncorrelated from component to component, is expected to scale as $N^{1/2}$. To maintain the presence of the noise in the large-$N$ limit, we scale the noise variance with the problem size, that is,
\begin{equation}
    \sigma_i^2 (N) = (\sigma^{\infty}_i)^2 \times N \ .
\end{equation}
Here, $(\sigma^{\infty}_i)^2$ is fixed as $N$ grows, so that signal and noise terms become comparable. The same argument applies to the other source of fluctuations in the data model, so that we have to choose
\begin{equation}
    \Delta_{t,t'} (N) = \Delta^{\infty}_{t,t'} \times N \quad \text{and} \quad 
     \Xi_{\tau,\tau'} (N) = \Xi^{\infty}_{\tau,\tau'} \times N \ .
\end{equation}
All observables are then obtained as functions of the rescaled parameters $\alpha, (\sigma^{\infty}_i)^2, \Delta^{\infty}_{t,t'} ,  \Xi^{\infty}_{\tau,\tau'} $, as well as of the other parameters that do not need to be rescaled, e.g. $F_\tau$, $x^{(k)}_{t}$, ... To lighten notations, we hereafter omit the $^\infty$ superscript in the rescaled parameters. However, it is important to keep in mind that, to compare to simulated data, the parameters $\sigma_i^2, \Delta ,  \Xi$ appearing in the expressions of the observables below should be replaced with their counterparts defined in Table~\ref{tab:data} and divided by the dimension of the data.

\subsection{Expression of the ground state energy and order parameters}

Our calculation gives the following expression for the ground-state energy:
 \begin{eqnarray}
		E_{GS} &=& \underset{v,\hat{v}, q,\hat q, \hat U, R, \hat{R}, W, \hat{W}, M, \hat M}{\text{optimum}} \left( -\frac{1}{2 N} \sum\limits_i \text{Tr} \left[\big( \bar{\sigma}_i^2\, \hat{W} + \hat U + \hat{v}^\dagger \, ({e}_i\otimes {e}_i )\big)^{-1} (\bar{\sigma}_i^2 \, \hat M + \hat q ^\dagger\,({e}_i\otimes {e}_i )  - \hat{R}\, {e}_i\,{e}_i^\dagger\, \hat{R}^\dagger ) \right] \right. \nonumber \\ 
        & + & \frac{\alpha}{T}  \text{Tr} \left[ H \,
        \left( M \otimes Z + q ^\dagger\, \mathcal{X} + R \, \text{diag}(\xi^2)\, R^\dagger  \otimes \bar{\Delta}\left(I_T -  J_T \right) \right)\right] \nonumber \\
        & + & \left. \frac{1}{2} \text{Tr} (\hat U) +  \frac{1}{2} \text{Tr}( \hat{v} ^\dagger q ) + \frac{1}{2}  \text{Tr}(\hat q^\dagger \, v)  + \text{Tr}(\hat{R}\, R^\dagger) + \frac{1}{2} \text{Tr}(\hat{W} \, M^\dagger) + \frac{1}{2} \text{Tr} (\hat M \, W^\dagger) \right) \ . \label{Before_saddle_point}
	\end{eqnarray}
    where $H$ is the $K.T\times K.T$-dimensional matrix defined through
    \begin{equation}\label{eq:defH}
        H=\left( I_{K.T} - 2\left(W \otimes Z + v^\dagger\,\mathcal{X}\right)\right)^{-1}\ .
    \end{equation}
In addition to the defining parameters of the model exposed in Section~\ref{sec:model}, the expression above involves several order parameters, whose definitions are reported in Table~\ref{tab:op}, as well as conjugated parameters, see list in Table~\ref{tab:cp}. Furthermore, we introduced the following notations:

\begin{table}
\begin{tabular}{|c|c|c|}
    \hline
    Notation & Definition & Number \\    \hline & & \\
    $ R^{(k,k')}$ & $\frac 1N \sum_i \big\lceil\langle v_i^{(k)}\rangle \big\rceil \, e_i^{(k')}$ & $K^2$ \\[.3cm]
    $ W^{(k,k')}$ & $\frac \beta N \sum_i \bar \sigma _i^2\big\lceil \langle v_i^{(k)}\, v_i^{(k')}\rangle- \langle v_i^{(k)}\rangle\, \langle v_i^{(k')}\rangle \big\rceil$ & $\frac 12 K(K+1)$ \\[.3cm]
 $ M^{(k,k')}$ & $\frac 1N \sum_i \bar \sigma _i^2\big\lceil \langle v_i^{(k)}\rangle\, \langle v_i^{(k')}\rangle\big\rceil$ & $\frac 12 K(K+1)$ \\[.3cm]
  $ v^{(k,k',\ell,\ell')}$ & $\frac \beta N \sum_i e_i^{(k)}e_i^{(k')}
  \big\lceil\langle v_i^{(\ell)}\, v_i^{(\ell')}\rangle -\langle v_i^{(\ell)}\rangle\, \langle v_i^{(\ell')}\rangle\big\rceil $ & $\big[\frac 12 K(K+1)\big]^2$ \\[.3cm]
  $ q^{(k,k',\ell,\ell')}$ & $\frac 1N \sum_i e_i^{(k)}e_i^{(k')}\big\lceil\langle v_i^{(\ell)}\rangle\, \langle v_i^{(\ell')}\rangle\big\rceil$ & $\big[\frac 12 K(K+1)\big]^2$ \\[.3cm]
    \hline
\end{tabular}
\caption{Order parameters appearing in the ground-state energy $E_{GS}$, see Eq.~\eqref{Before_saddle_point}.}
  \label{tab:op}
\end{table}

\begin{table}
\begin{tabular}{|c|c|c|}
    \hline
    Notation & Conjugated to & Number \\
    \hline & & \\
    $ \hat R^{(k,k')}$ & $ R^{(k,k')}$ & $K^2$ \\ [.3cm]
    $ \hat W^{(k,k')}$ & $W^{(k,k')}$ & $\frac 12 K(K+1)$ \\[.3cm]
 $ \hat M^{(k,k')}$ & $M^{(k,k')}$ & $\frac 12 K(K+1)$ \\[.3cm]
  $ \hat v^{(k,k',\ell,\ell')}$ & $v^{(k,k',\ell,\ell')} $ & $\big[\frac 12 K(K+1)\big]^2$ \\[.3cm]
  $ \hat q^{(k,k',\ell,\ell')}$ & $q^{(k,k',\ell,\ell')}$ & $\big[\frac 12 K(K+1)\big]^2$ \\[.3cm]
  $\hat U^{(k,k')} $ & normalization and orthogonality & $\frac 12 K(K+1)$\\
   & of $\{\mathbf{v}^{(k)}, k=1,...,K\}$ & \\[.3cm]

    \hline
\end{tabular}
\caption{Conjugated parameters appearing in the ground-state energy $E_{GS}$, see Eq.~\eqref{Before_saddle_point}.}
  \label{tab:cp}
\end{table}

\begin{itemize}
\item the symbol $\otimes $ for tensor product, e.g. $W\otimes Z$ is a $K.T\times K.T$-dimensional matrix with entries $(W\otimes Z)^{(k,k')}_{t,t'} =W^{(k,k')}\times Z_{t,t'} $ and $e_i \otimes e_i$ is a $K^2$-dimensional vector with elements $(e_i \otimes e_i) ^{(k,k')}=e_i^{(k)}e_i^{(k')}$.
\item $q$, $v$, $\hat q$, $\hat v$ are considered as $K^2  \times K^2$ matrices, in which rows and columns are labeled, respectively, by pairs of integers $(k,k')$ and $(\ell,\ell')$. They are built by symmetrizing the 4-tensors defined in Tables~\ref{tab:op} and \ref{tab:cp}.  
         \item Tr denotes the trace operator, e.g. Tr$(\hat U)= \sum _{k} \hat U^{(k,k)}$. We also define the partial trace over time indices for matrices $A$ of dimension $K.T$: Tr$_T( A)= \sum _{t}  A^{(k,k')}_{t,t}$, hence defining a $K\times K$ matrix.
        \item $I_m$ is the identity matrix of size $m\times m$.
        \item $J_m$ is the uniform matrix of size $m\times m$, whose all entries are equal to $\frac 1m$.
        \item $^\dagger$ denotes the matrix transpose, e.g. $(M^\dagger)^{(k,k')}= M^{(k',k)}$.
   \end{itemize}

\subsection{Resolution of the saddle-point equations}

To determine the order parameters and the conjugated parameters, we look for a saddle point of the ground-state energy. In practice, we search  for solutions of the equations
\begin{equation}
    \frac{\partial E_{GS}}{\partial v} = 0,~~~\frac{\partial E_{GS}}{\partial \hat{v}} = 0, ~~~\dots
\end{equation}

The vanishing derivative conditions may correspond to either minimization or maximization with respect to the corresponding parameters. These two possibilities are consequences of two features of our calculation:

\vskip .3cm
\noindent {\bf Complex Integration Over Conjugated Parameters.} The integrals over the Lagrange multipliers conjugated to the "physical" order parameters in Table~\ref{tab:cp} run over the imaginary axis, as imposed by the integral representation of the Dirac distributions. The integration contours are then deformed in the complex domain to go through the saddle points, which lie on the real axis; the second order derivative at the saddle-point along the real axis has opposite sign to the one along the imaginary direction. 

\vskip .3cm
\noindent {\bf Replica peculiarities.} Our calculation relies on the replica method, in which $n\to 0$ copies of the system (here, a set of $K$ vectors) are interacting after integrating out the data \cite{Legenkaia2024}. As a result of the analytic continuation, terms in $E_{GS}$ that involve $\frac12 {n(n-1)}$ pairs of replicas must be maximized rather than minimized when $n<1$ \cite{Steinberg2024}. 

\vskip .3cm
Notice that the expression for $E_{GS}$ is linear in $M$ and in $q$, entailing that $\hat W$ and $\hat v$ can be readily expressed in terms of the other parameters. However, eliminating these parameters would require substantial matrix inversion, and we found it more convenient from a numerical point of view to optimize the entire $E_{GS}$ function. 

\subsubsection{Second-order method for saddle-point determination} 

The energy $E_{GS}$ depends on $K^4+2K^3+\frac {11}{2} K^2+\frac 52 K$ parameters, see Tables~\ref{tab:op} and \ref{tab:cp}. Search for the saddle-point in this space can be seen as a maximization/minimization problem, depending on the direction considered. To simplify and automatize the extremization procedure, we formally collect all parameters into a single  vector $\omega$. We then use Newton's method \cite{newton}. The key steps of this second-order procedure are:

\begin{enumerate}
    \item {\em Gradient calculation.} At each iteration, we compute the gradient $\nabla E_{GS}$  of the energy with respect to the parameters. This gradient gives us the direction of the steepest ascent or descent for each parameter.
    \item {\em Second derivatives calculation.} We then compute the Hessian matrix, $\mathcal {H}_{GS}$, which is the matrix of second derivatives of the energy with respect to each pair of parameters. The Hessian encodes information about the local curvature of the energy.
    \item {\em Update step.}  
    Once we have the gradient and Hessian, Newton’s method updates the order parameters according to the rule:
    \begin{equation}
        \omega_{\text{new}} = \omega_{\text{old}} - \big(\mathcal{H}_{GS}\big)^{-1}\, \nabla E_{GS} \ .
    \end{equation}
    This update step guarantees that (1) the energy is appropriately maximized or minimized (depending on the sign of the eigenvalues of $\mathcal{H}_{GS}$), (2) the amplitude of the step size adapts to  the local curvature—taking smaller steps in directions with large curvature and larger steps where the curvature is lower. This allows us to move efficiently towards the saddle point.
\end{enumerate}





\subsubsection{Starting point for the resolution procedure.} 

To initialize Newton's procedure, we consider the ideal case of no noise ($\sigma_i^2=0$) and no sample-dependent signal fluctuation ($\Xi=\mathcal{X}=0$). 
In this case, the vanishing conditions for the derivatives of $E_{GS}$ with respect to $\hat W$ and $\hat M$ yield $M=W=0$. This result is expected from the definition of $M$ and $W$ in Table~\ref{tab:op}. Similarly, differentiating with respect to
$q$ and $v$ implies that their conjugated forces vanish: $\hat q= \hat{v} = 0$.

Let us now consider the $K\times K$--matrix $A=\text{diag}(\xi^2)\otimes\frac 1T\text{Tr}\big(\bar \Delta(I_T- J_T)\big)+\frac 1T \bar x^\dagger\, \bar x$.
We write $A=P^\dagger\, D\, P$, where $D$ denotes the diagonal matrix made with the eigenvalues of $A$, and $P$ the orthogonal matrix collecting its eigenvectors. Then, setting the derivatives with respect to $\hat R, R, \hat U$ to zero, we obtain $R=P$, $\hat R=-2\alpha\, D\, P$, and $\hat U=2\alpha\, D$. Notice that $P=I_K$ in the absence of sample-to-sample fluctuations ($\xi=0$) and of smoothing, since the $x_t^{(k)}$ vectors are orthogonal. Upon substitution into Eq.~(\ref{Before_saddle_point}) and taking the derivatives with respect to $M,W,\hat{v},\hat q$, we  find the remaining order and conjugated parameters: 
\begin{eqnarray}
q=\frac 1N \sum_i e_i \, e_i^\dagger\, (P\, e_i)\, (P\, e_i)^\dagger\ , \quad v=\frac 1{2\alpha}\, I_K \otimes D^{-1} \ ,\quad \hat W=-2\alpha\, I_K\ ,\nonumber \\
\hat M= -4\alpha \, P\, \text{diag}(\xi^2) \, P^\dagger \times \frac 1T \, \text{Tr} \left( Z \, \bar \Delta \, \big( I_T - J_T\big) \right)
-4\alpha\, P \left( \frac 1T\, \bar x^\dagger Z\, \bar x\right) P^\dagger  
\ . 
\end{eqnarray}

To find the saddle-point solution for arbitrary $\bar{\sigma}_i$ and $\bar{\Xi}$, we start from the exact solution above valid for  $\bar{\sigma}_i=0$ and $\bar{\Xi}=0$, and gradually introduce the variability in the measurement convolution kernels by increasing $\bar{\Xi}$, while tracking the saddle-point at the same time. We then repeat this procedure by increasing $\bar{\sigma}_i$ from 0 to its target value.

\subsection{Expressions of observables}
\label{sec:observ}

Once the order parameters are determined, we obtain the $\epsilon$ matrix through

   \begin{eqnarray}
        \epsilon  &=& \frac{1}{N}  \,\bar{x}^\dagger\, \bar{x} -  \frac{2}{N}\,  \text{Tr}_T \big[ H\, (R \otimes I_T) \, X \big] + \frac{1}{N}  \text{Tr}_T \big[ H\,\left( M \otimes Z + q^\dagger\, \mathcal{X} + R\, \text{diag}(\bar{\xi}^2)\, R^\dagger \otimes \bar{\Delta}\, J_T \right) \big]  \nonumber \\
        & + & \frac{1}{N} \text{Tr}_T \big[ H \, \left( M \otimes Z + q^\dagger\,  \mathcal{X} + (R \otimes I_T) (  \text{diag}(\bar{\xi}^2) \otimes \bar{\Delta} \big(I_T - J_T\big )+ X) (R^\dagger \otimes I_T) \right) H \big] \ ,
    \end{eqnarray}
where $H$ is defined in Eq.~\eqref{eq:defH} and $X$ is the $K.T\times K.T$-dimensional matrix with entries $X_{(k,t),(k',t')}=\bar x_t^{(k)} \,\bar x_{t'}^{(k')}$.

The expression for the $\rho$ matrix is immediately obtained from the definition in Eq.~\eqref{eq:defrho}:
\begin{eqnarray}
 \rho = I_{K} - \frac 12 \big( R+R^\dagger\big) \ .
 \label{rho_definition}
\end{eqnarray}
We have also derived a local measure of the uncertainty of each component $i$, 
  $\rho^{(k,k')} _i = \frac{1}{2}\langle ( v^{(k)}_i  - e^{(k)}_i )( v^{(k')}_i  - e^{(k')}_i )\rangle$, with the result
\begin{eqnarray}\label{def:rhoi}
	\rho_i  &=&  -\frac{1}{2 } \big( \bar{\sigma}_i^2 \,\hat{W} + \hat U + \hat{v}^\dagger\, e_i\,  e_i^\dagger \big)^{-1} \big(\bar{\sigma}_i^2 \, \hat M + q^\dagger\, e_i \,e_i^\dagger - \hat{R} \, e_i\, {e}_i^\dagger \,\hat{R}^\dagger \big) 
    \big( \bar{\sigma}_i^2 \,\hat{W} + \hat U + \hat{v}^\dagger\, e_i\,  e_i^\dagger \big)^{-1}
    + \nonumber \\
        & + & \frac{1}{2 } \, {e}_i\, {e}_i^\dagger  +   \frac 12 \big( \bar{\sigma}_i^2 \,\hat{W} + \hat U + \hat{v}^\dagger\, e_i\,  e_i^\dagger \big)^{-1}  ( \hat{R}+\hat{R}^\dagger ) \, {e}_i \,{e}_i^\dagger \ .
	\end{eqnarray}  
It can be explicitly checked, using the saddle-point equations for $E_{GS}$, that $\rho$ in Eq.~\eqref{rho_definition} is recovered by summing $\rho_i$ above over all components $i$ and dividing by $N$ \cite{Legenkaia2024}.

\section{Validation on synthetic data}


To test our predictions, we explore a range of values for the parameters of the data model, varying one parameter at a time while holding the others constant. For each set of parameter values, we generate synthetic data,  carry out PCA, then estimate $\rho$ and $\epsilon$. These numerical estimates are compared with the analytical predictions obtained in Section~\ref{sec:calcul}. 

\subsection{Tests in the absence of measurement convolution}

For the sake of simplicity, we first consider a data model with no measurement convolution $(F_{i,\tau}=\delta_{\tau,0})$:

\begin{itemize}
    \item We consider $K=2$-dimensional latent activity and set the temporal profile of the signal in the following way:
    \begin{equation}\label{eq:signal2}
        \begin{cases}
            x^{(1)}_t = a^{(1)} \left( \sin\left(\frac{2\pi t}{T} \right) + \sin\left(\frac{4 \pi t}{T} \right) \right) \\
            x^{(2)}_t = a^{(2)} \left( -\sin\left(\frac{2\pi t}{T} \right) + \sin\left(\frac{4 \pi t}{T} \right) \right)
        \end{cases}
    \end{equation}
    where the coefficients $a^{(1)}$ and $a^{(2)}$ fix the magnitude of the signal.
    \item The modes $e^{(1)}$ and $e^{(2)}$ are randomly chosen as two orthogonal vectors from the uniform distribution over $N$-dimensional normalized vectors.
    \item For the sample-to-sample variations $\delta x^{(k)}$, we choose the temporal correlation matrix $\Delta$ to be Gaussian, with correlation width $\tau_{\xi}$:
    \begin{equation}
        \Delta_{t_1,t_2} = e^{-(t_1-t_2)^2/ (2 \tau_{\xi}^2)}\ .
    \end{equation}
    \item The standard deviations $\xi^{(k)}$ of the sample-to-sample fluctuations are chosen to be proportional to the amplitudes $a^{(k)}$ of the signal components, so that 
    \begin{equation}
        \frac{\xi^{(1)}}{a^{(1)}} = \frac{\xi^{(2)}}{a^{(2)}}\equiv  \xi\ .
    \end{equation}
    \item We choose the variance of the noise $z$ to be uniform across all neurons $i$, {\em i.e.} $\sigma_i=\sigma$.
    \item The kernel $G$ used for data smoothing is Gaussian, with width $\tau_z$.
    
\end{itemize}

\subsubsection{Results for $\epsilon$}

\begin{figure}[bth!]
{\includegraphics[width=\linewidth]{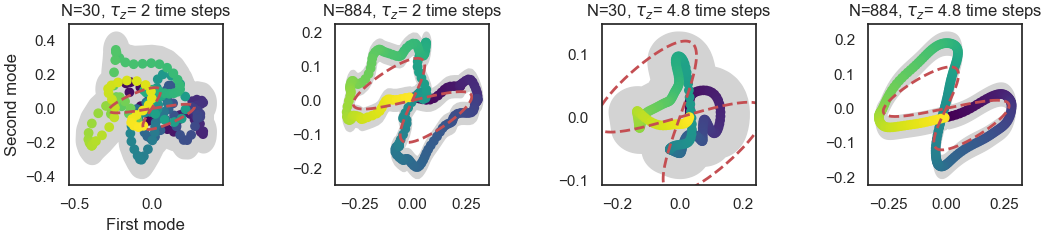}}
	\caption{Example trajectories reconstructed after application of PCA to synthetic data. The ground-truth latent dynamics is given by the two-dimensional tilted $\infty$-shape, see Eq.~\eqref{eq:signal2}, shown with the dashed red line. The quality of the inference depends of the parameters of the dataset. The inferred trajectories are shown as a gradient from the first  (purple) to the last (yellow) time step. Gray color identifies the confidence area around the reconstructed trajectories, with a width given by $\sqrt{\epsilon}$, see text. Parameter values: $N$ and $\tau_z$ are reported above the panels, $T=200$, $a^{(1)}=0.1$, $a^{(2)}=0.045$, $\sigma_i =\sigma=1$, and $\xi=0.017$.
    }
    \label{fig:epsilon1}
\end{figure}

We show in Figure \ref{fig:epsilon1} example reconstructions of the signal trajectories for different numbers $N$ of probed components and widths $\tau_z$ of the smoothing kernel. We observe an improvement of the quality of the inferred trajectory for larger $N$ and $\tau_z$


\begin{figure}[bth!]	{\includegraphics[width=\linewidth]{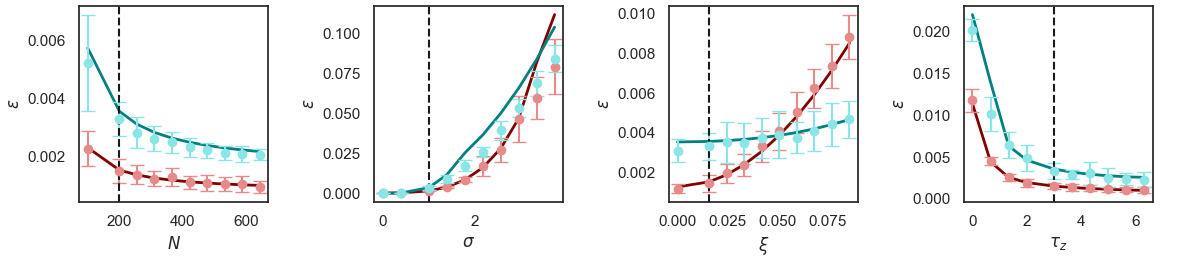}}
	\caption{Comparison of analytical predictions and results of PCA on synthetic data with the two-dimensional tilted $\infty$-shaped latent dynamics when (from left to right) $N$, $\sigma$, $\xi$ and $\tau_z$ are varied. $\epsilon^{(1,1)}$ is shown in red, $\epsilon^{(2,2)}$ in cyan. Theoretical predictions are shown as solid lines, average values obtained from generated data are shown with error bars. Across all plots, dataset parameters are varied around a reference point, indicated by a dashed line: $T=N=200$, $a^{(1)}=0.1$, $a^{(2)}=0.045$, $\sigma_i =1$, $\tau_z=3$, and $\xi=0.017$.
    }
    	\label{fig:epsilon2}
\end{figure}

However, even with infinitely large number of components, achieving perfect accuracy is impossible, {\em i.e.} $\epsilon$ does not vanish for $N\to\infty$, see Figure~\ref{fig:epsilon2}. The underlying intuition behind this limitation lies in the increasing dimensionality of the problem. As $N$ grows, so does the number of components that define the modes to be inferred. Informally speaking, while increasing $N$ should {\em a priori} provide more information about the signal trajectory, this information is conveyed in an increasingly complex manner.

We then study in Figure \ref{fig:epsilon2} how the accuracy on the trajectory depends on the measurement noise. When $\sigma$  vanishes, the trajectory is perfectly recovered, despite the presence of sample--to--sample variability.  Conversely, as the measurement noise increases, so does $\epsilon^2$, with an abrupt change of slope at high values of $\sigma$. We will comment on this phenomenon below.

As expected, when all other parameters are fixed, increasing sample--to--sample variability results in an increase of the error $\epsilon$. Interestingly, as $\xi$ increases the error on the reconstruction of the latent trajectory may become smaller for the second mode than for the first one, though the signal coordinate show weaker variability for the former than for the latter. The reason for this apparently counterintuitive phenomenon is that adding sample-to-sample variability has a two-fold effect on performance. On the one hand, it enhances the fluctuations along the modes and thus helps for their recovery. On the other hand, it increases the dispersion around the mean latent signal and is therefore detrimental for the accuracy of the trajectory reconstruction. These two effects contribute to $\epsilon$ in different ways depending on the precise setting, and may lead to the crossing observed in the figure.

Similarly, averaging the data with a wider smoothing kernel, {\em i.e.} larger $\tau_z$ improves accuracy. More data points are averaged together, making it easier to identify the underlying signal. However, we expect that increasing $\tau_z$ beyond the characteristic auto-correlation time of the signal, which is scaling linearly with $T$ in the example considered here, see Eq.~\eqref{eq:signal2}, would wash away important features and degrade performances.

\subsubsection{Results for $\rho$}
\label{sec:descriABCD}

\begin{figure}[bth!]
{\includegraphics[width=\linewidth]{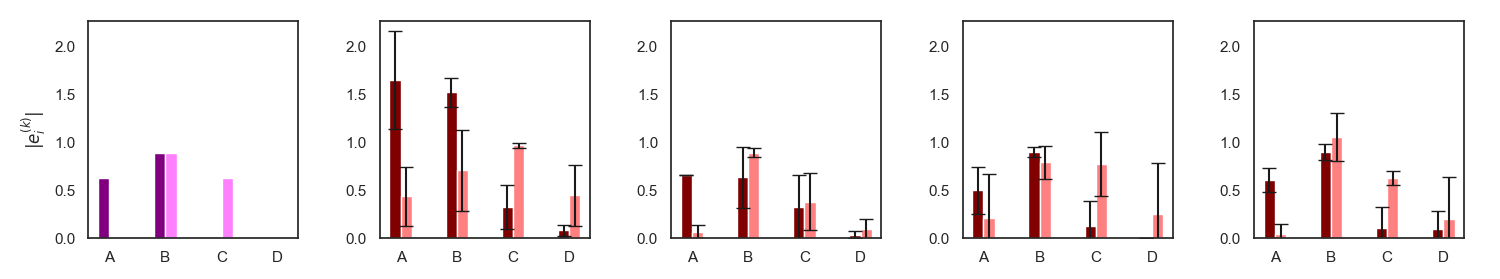}}
	\caption{{Left:} Components of the latent modes $e^{(k)}_i$ for four different components $i$ denoted by A,B,C,D in the synthetic data. The first mode $e^{(1)}$ is shown in dark purple, and the  second one, $e^{(2)}$, in pink. {Rest of the row:} Examples of $v^{(k)}_i$ for $i=$A,B,C,D and for synthetic data with different sizes $N$ and smoothing kernel width $\tau_z$, same values as in Figure~\ref{fig:epsilon1}. The first mode, $v^{(1)}$, is shown in dark red, and the second, $v^{(2)}$,  in light red. Error bars are the squared roots of the accuracies $\rho^{(k,k)}$ in Eq.~\eqref{eq:defrho}.
    }
    \label{fig:rho1}
\end{figure}

We show in Figure~\ref{fig:rho1} how four components $i$, hereafter referred to as $A$, $B$, $C$ and $D$ for simplicity contribute to the two modes. $A$ and $C$ are `pure' components participating in, respectively, modes 1 and 2 only. $B$ is a mixed component with equal contributions, {\em i.e.} $e^{(1)}_B=e^{(2)}_B$. $D$ is not part of either mode. We then show, for the same values of $N$ and $\tau_z$ as in Figure~\ref{fig:epsilon1}, the inferred projection $v^{(k)}$ for these four representative cases. The error bars, estimated as $\sqrt{\rho^{(k,k)}}$, allow one to assess whether the projections are reliably different from zero.

For low $N$ and $\tau_z$, the inferred components largely differ from the ground truth values: $A$ and $C$ appear to be mixed, $B$ seems to have different amplitudes on the two modes, and $D$ is apparently part of mode 2. As $N$ increases, inference improves,  but inference is still misleading for some components, {\em e.g.} the mode amplitudes appear to differ for component $C$ in the third panel of Figure~\ref{fig:rho1}. Using wider smoothing kernels helps recovering the qualitative behavior of the different latent mode components. We stress that the ground-truth mode components are always compatible with their inferred counterparts, even at low $N$ and $\tau_z$ where the inference quality is poor, when error bars are taken into account. This confirms the reliability of our calculation of $\rho$ and its usefulness to assess the quality of  mode reconstruction.

\begin{figure}[bth!]	{\includegraphics[width=\linewidth]{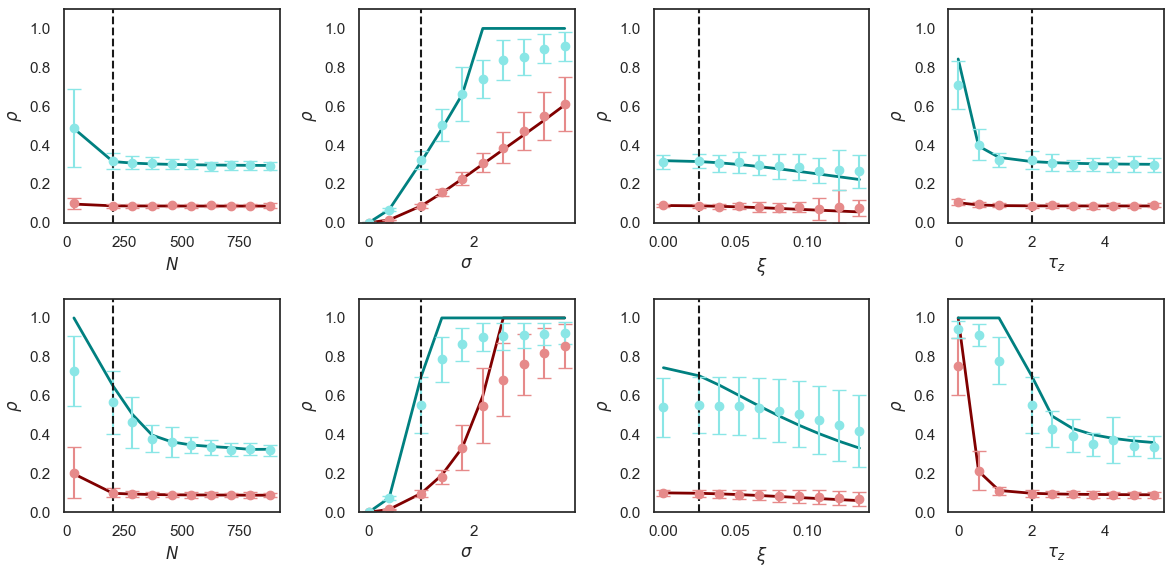}}
	\caption{Comparison of the prediction and the results of PCA performed on synthetic data with the two-dimensional tilted $\infty$-shaped latent dynamics. $\rho^{(1,1)}$ is shown in red, $\rho^{(2,2)}$ in cyan. 
    Top row: case of homogeneous measurement noise across all components, $\sigma_i=\sigma$.
    Bottom row: case of heterogeneous measurement noise,  $
    \sigma_i$ were randomly drawn from a centered Gaussian distribution, and then squared and rescaled such that the mean value is $\sigma^2$.
    Theoretical predictions are shown as solid lines, average values obtained from generated data are shown with error bars. Across all plots, dataset parameters are varied around a reference point, indicated by a dashed line, see Figure~\ref{fig:epsilon2}.
    }
    \label{fig:rho2}
\end{figure}

We then systematically investigate the behavior of the reconstruction error $\rho$ of the signal modes. As already commented above, increasing the number of probed components does not guarantee perfect recovery of the modes: $\rho$ saturates to a nonzero value as $N\to\infty$, see Figure~\ref{fig:rho2}, left panel. The asymptotic value of $\rho$ is, however, lower for signal modes having larger amplitude.

The accuracy $\rho$ is, as expected, an increasing function of the measurement noise level. A phase transition is encountered at a critical value of $\sigma$, above which $\rho=1$, see Figure~\ref{fig:rho2} and the inferred modes $v^{(k)}$ are not at all aligned along the ground-truth modes $e^{(k)}$. This phenomenon is reminiscent of the retarded learning phase transition taking place in the spiked covariance model, and is associated to the cusp in the representative curve of $\epsilon$ shown in Figure~\ref{fig:epsilon2}. The presence of the no-recovery phase is also observed when the width of the smoothing kernel is too small, and the measurement noise is not sufficiently averaged out.

\begin{figure}[bth!]	{\includegraphics[width=.3\linewidth]{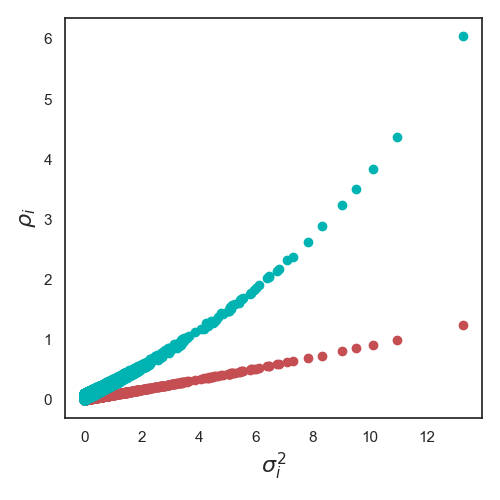}}
	\caption{Dependence of the accuracy $\rho_i$ in reconstructing component $i$ of the latent modes, see Eq.~\eqref{def:rhoi}, vs. local measurement noise variance $\sigma_i^2$. Results for the first mode are shown in red, and in cyan for the second mode. All parameters are fixed to the values corresponding to the dashed vertical lines in Figure~\ref{fig:rho2}(bottom).
    }
    \label{fig:rho3}
\end{figure}

We also observe that heterogeneity in the measurement noise strongly affects the quality in the mode recovery. More precisely, keeping the same global level of noise, {\em i.e.} $\sum_i \sigma_i^2$, the components $i$ plagued by large noise are harder to recover, as shown in Figure~\ref{fig:rho3}. This phenomenon leads to a global increase of $
\rho$, especially for low-variance modes, compare top and bottom panels in Figure~\ref{fig:rho2}.

Interestingly, the presence of sample-to sample variability, which is deleterious for the reconstruction of the signal trajectory, is beneficial for the inference of the modes, see the decreasing behavior of $\rho$ with $\xi$ in Figure~\ref{fig:rho2}.  This can be understood as follows. While $\delta x^{(k)}$ is a noise term, it occurs along the direction $e^{(k)}$, as the signal $x^{(k)}_t$. As a result, it contributes to increasing the variance along the direction $e^{(k)}$, making it easier to infer.

\begin{figure}[bth!]	{\includegraphics[width=\linewidth]{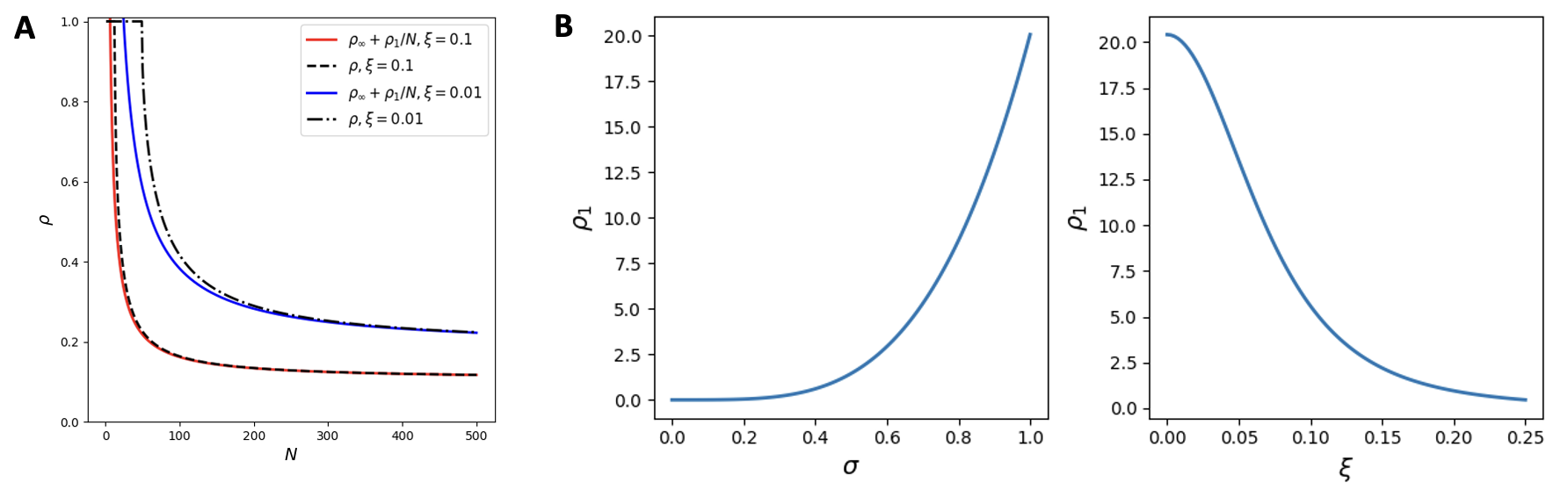}}
	\caption{Finite-size effects. {\bf A.} Accuracy $\rho$ vs. number $N$ of measured components for two values of the sample-to-sample fluctuations, $\xi=0.01$ (top, dashed-dotted line) and $\xi=0.1$ (bottom, dashed line). The full lines show the approximation defined in Eq.~\eqref{eq:approxN}. {\bf B.} Dependence of $\rho_1$ on $\sigma$ (left) and $\xi$ (right). Parameter values: $K=1$, $T=200$, $\text{var}(x) = 0.01$, $\sigma = 1$ for the left and right panels, and $\xi=0.01$ for the middle panel.
    }
    \label{fig:rho4}
\end{figure}

As can be observed in Figure  \ref{fig:rho2}, $\rho$ decays with $N$, and reachs a limit value as the number of components tends to infinity\footnote{The same behaviour can be observed for $\epsilon$, see Figure~\ref{fig:epsilon2}.}. We hereafter study the rate at which this convergence takes place. To get analytically tractable formulas, we focus on the case of a single latent mode ($K=1$), homogeneous noise ($\sigma_i=\sigma$), no smoothing kernel, and we discard any time correlations in the sample-sample fluctuations, {\em i.e.} $\Delta =I_T$. With this simple setting, the overlap between the inferred and ground-truth mode can be analytically calculated, with the result
\begin{equation}
   R = \sqrt{\frac{\big(\text{var}(x)+ \xi^2\big)^2\, T- \frac{\sigma^4}{N}}{\big(\text{var}(x) + \xi^2\big) \big( (\text{var}(x)+\xi^2) \,T+ \sigma^2\big) }} 
\end{equation}
Using $\rho = 1-R$ according to Eq.~\eqref{rho_definition} and expanding in powers of $\frac 1N$, we get
\begin{equation}\label{eq:approxN}
\rho=\rho_\infty+\frac{\rho_1}N \ ,
\end{equation}
with
\begin{eqnarray}
    \rho_{\infty} = 1 - \sqrt{\frac{\big(\text{var}(x) + \xi^2\big) \,T}{ \big(\text{var}(x) + \xi^2\big) \, T+ \sigma^2 }},~~~~ \rho_1 = \frac{\sigma^4 }{2 \sqrt{T \big(\text{var}(x) + \xi^2\big)^3\, \big( (\text{var}(x) + \xi^2) \,T+\sigma^2\big)}}\ .
\end{eqnarray}
We  show in Figure~\ref{fig:rho4}A the behavior of $\rho$ as a function of $N$ for two values of $\xi$, while the other parameters are kept fixed. For comparison, we also plot the predictions obtained  with formula \eqref{eq:approxN}, which neglects $\frac1{N^2}$ and higher-order corrections. We see that the agreement is very good, as soon as $N$ exceeds $\rho_1$.

The coefficient controlling finite-$N$ corrections, $\rho_1$, is plotted as function of control parameters in Figure~\ref{fig:rho4}B. Lower noise or larger signal fluctuations not only make the asymptotic value $\rho_\infty$ smaller, but also decrease $\rho_1$. In other words, the convergence towards the limit value is faster (with $N$) when the signal is stronger.

\subsection{Tests on synthetic data with measurement convolution kernels}

We now test the validity of our predictions, in the presence of measurement convolution, {\em i.e.} $F\ne 0$, and of fluctuations in those kernels, {\em i.e.} $\Xi\ne 0$. We generate the convolution kernels as follows:
\begin{equation}
    F_{i,\tau}=\exp(-\tau/d_i)-\exp(-\tau/r_i) \ ,
\end{equation}
where $r_i$ and $d_i$ are, respectively, the rise and decay times of the kernel (with $r_i\ll d_i$). In practice, we draw these rise and decay times from two Gamma distributions $\Gamma (\cdot;k,\theta)$ with two distinct sets of shape  and scale parameters, which we denote by, respectively, $(k_r,\theta_r)$ and $(k_d,\theta_d)$. We stress that this choice, while reasonable from a physical point of view, is arbitrary and that many other shapes for $F$ could be considered.

\begin{figure}
    \centering
\includegraphics[width=0.8\linewidth]{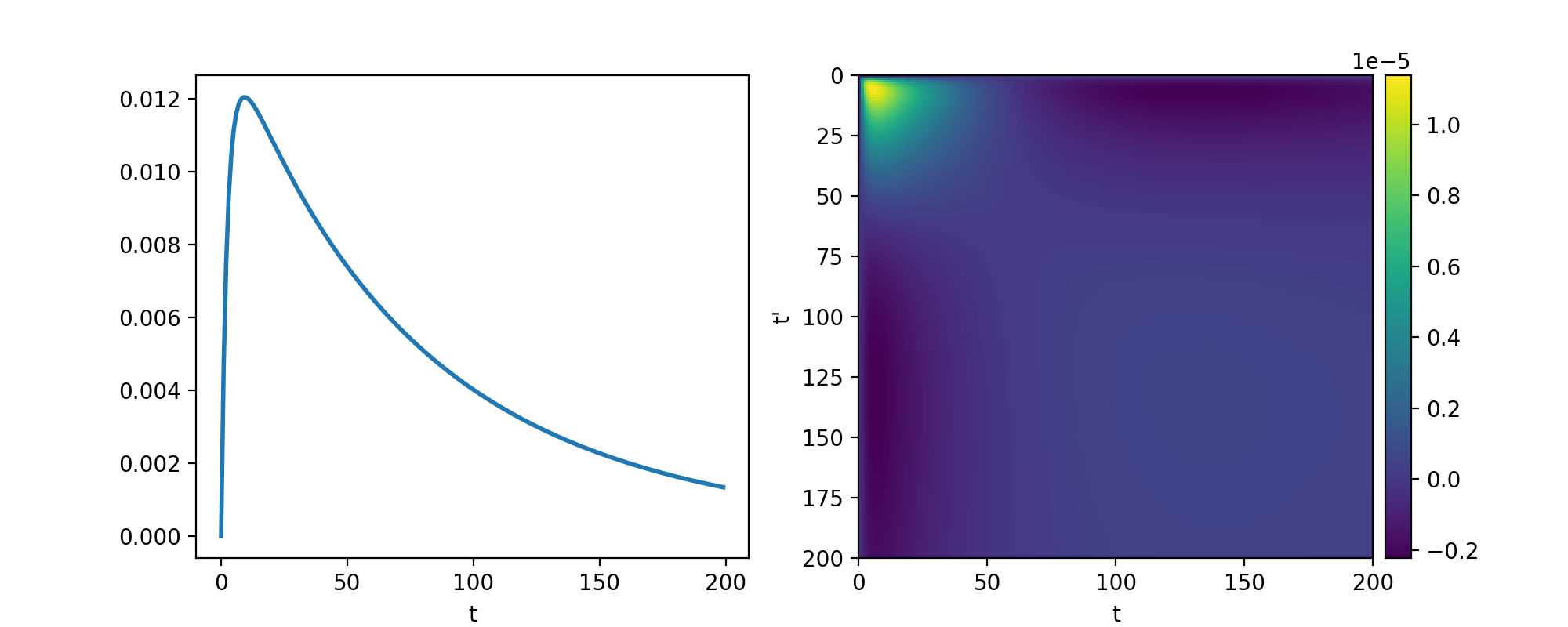}
    \caption{Case of measurement convolution kernel, see text for parameter values. Left: Mean  kernel $ F_t$. Right: Correlation matrix $\Xi_{t,t'}$ of the fluctuations $\delta F$, see Eq.~\eqref{eq:defxi}.
    }  \label{fig:kernel1}
\end{figure}

Hereafter we arbitrarily choose $(k_r,\theta_r)=(12.25,0.43)$ and $(k_d,\theta_d)=(2.78,28.8)$; other choices of parameters were considered, leading to qualitatively similar results, and are not shown.
We display in Figure~\ref{fig:kernel1}(left) the resulting average kernel $F_\tau$, with a steep rise at low delay $\tau$, followed by a longer decay at large delays. 
In Figure~\ref{fig:kernel1}(right) we represent the covariance matrix $\Xi$ of the kernel fluctuations $\delta F _\tau=F_{i,\tau}-F_\tau$. 

\begin{figure}
    \centering
\includegraphics[width=1\linewidth]{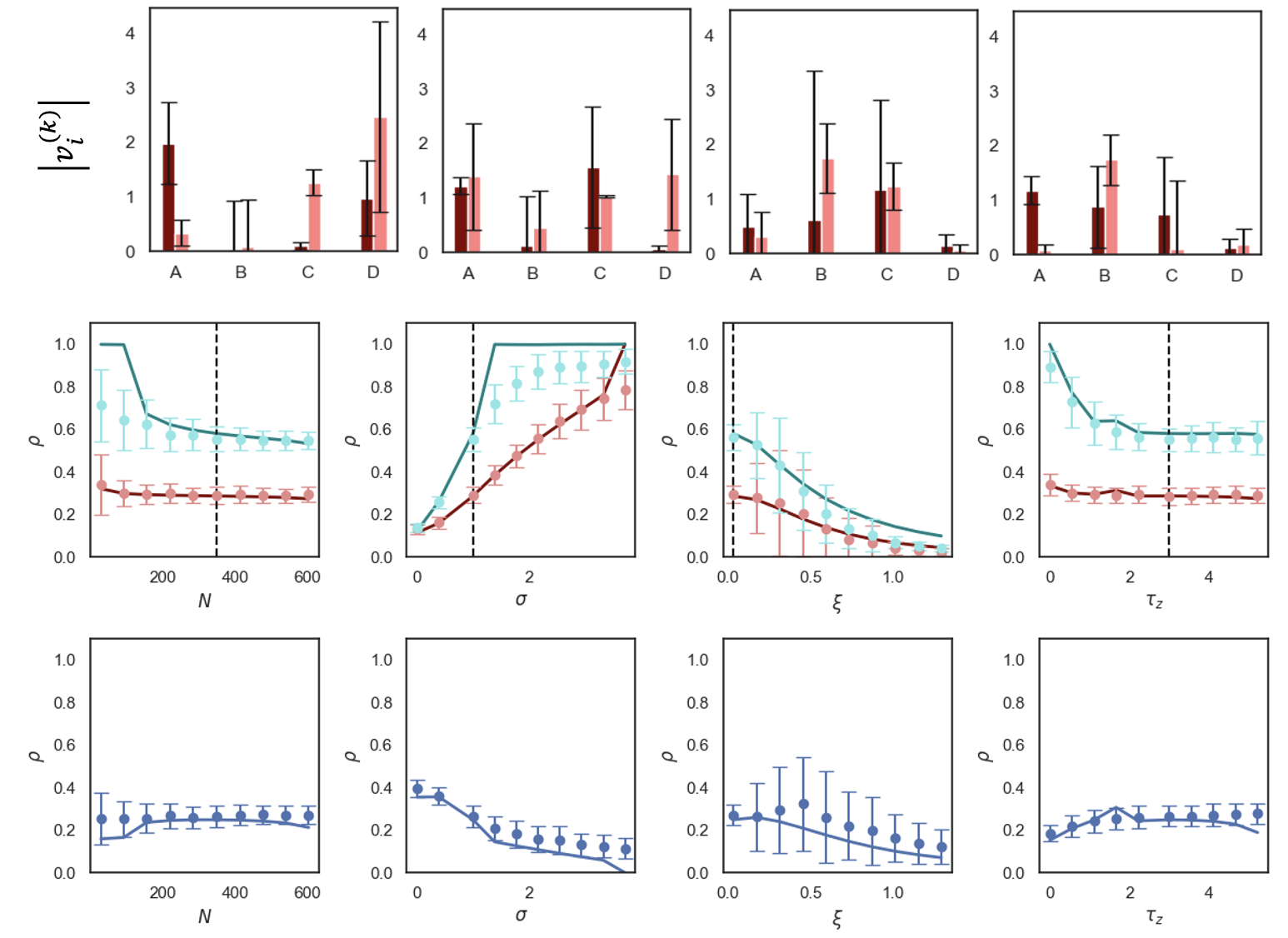}
    \caption{ Effects of measurement convolution kernels on the quality of mode reconstruction. Top: Components  $v^{(k)}_i$ of the inferred latent modes $k=1,2$, for $i=A,B,C,D$; the ground truth components $e^{(k)}_i$ are shown in the leftmost panel of Figure~\ref{fig:rho1}. Values of $N$ and $\tau_z$ are the same as in Figure~\ref{fig:epsilon1}. Middle: Diagonal elements of the accuracy matrix, $\rho^{(1,1)}$  in red and $\rho^{(2,2)}$ in cyan. Bottom: Off-diagonal element $\rho^{(1,2)}$ of the accuracy matrix.
    }
    \label{fig:kernel2}
\end{figure}

The presence of the convolution kernel does not affect the overall trends in the dependencies of $\epsilon$ and $\rho$ on the different parameters of the model that we have observed in the previous section, as reported in Figure~\ref{fig:kernel2}. As expected, accuracy still improves with the number $N$ of components, decreases with the noise strength $\sigma$, and benefits from wide smoothing kernels. The sample-to-sample fluctuations $\delta x$ remains beneficial for the recovery of the right directions of the modes, but makes the  recovery of the latent trajectory worse.

However, the existence of component-dependent fluctuations in the rise and decay times of the kernel produce several effects not observed for the model in the absence of measurement convolution. Notably, the recovery of the direction of the principal component can never be perfect, as seen by the residual value of $ \rho >0$ even in the absence of the measurement noise, see results for $\sigma=0$ in Figure~\ref{fig:kernel2}. We stress that this effect is not due to the presence of the (mean) kernel $F$, but to the one of the fluctuation $\delta F_i$ on the component $i$. In fact, larger variations in the rise and decay times lead to stronger values for $\rho$ at vanishing noise (not shown).

The presence of the kernel also introduces off-diagonal terms for $\rho$ even for very weak measurement noise, contrary to the no-kernel case. The fluctuations in the kernel, which vary with the component $i$, act as multiplicative factors onto every mode components $e_i^{(k)}$. As a result, the ground-truth directions of the latent modes cannot be inferred any longer. The off-diagonal terms in $\rho$ expresses how much the reconstructed modes are mixed up with respect to their ground-truth counterparts, see Eq.~\eqref{eq:defrho}. We observe in Figure~\ref{fig:kernel2}(bottom) that the confusion between the two modes is large even when the measurement noise $\sigma$ vanishes. As the measurement noise increases, the two modes are very poorly recovered, as shown from the increase in the diagonal elements $\rho^{(k,k)}$, while the off-diagonal element decreases. In this regime, the  latent modes are inferred very unreliably, and the large errors in their reconstruction are simply uncorrelated between components.

\section{Conclusion} 

In this work, we have derived analytical expressions for the accuracy of reconstruction of low-dimensional latent dynamics and space from high-dimensional data. Our framework takes into account a variety of measurement features, which are often encountered in real context. First, we allow for the measurement noise to vary from component to component. Second, we consider the effects induced by non-instantaneous measurements: the signal associated to a component is convoluted by a kernel, whose shape may change from component to component. Third, we acknowledge that complex systems' dynamics may be intrinsically noisy: repeating the same experiment may result in variable behavior, a fact that we take into account with sample-to-sample fluctuations. Our data model is then an extension of the so-called spiked covariance model, which incorporates the presence of these three sources of contributions to the measurements. 

Based on statistical physics methods, we were able to calculate analytically the accuracy of the reconstruction of the signal trajectories and of the modes as functions of the model-defining parameters. From a technical point of view, we used the replica approach, which is quite versatile, as it allowed us to deal with our involved data model. Let us stress that we have resorted to the so-called replica symmetric Ansatz, as we do not expect any breaking of symmetry in the present setting: the ground state of the energy over the $\mathbf{v}^{(k)}$'s in Eq.~\eqref{eq:energy} is unique, up to an arbitrary permutation of the $K$ vectors. 

As in the simpler spike covariance model, we observe the onset of a phase transition as the level of measurement (or sampling) noise decreases, characterized by the emergence of non-zero overlaps $R$ between the ground-truth and inferred mode vectors.  The location of this transition depends not only on the amplitude of the signal along the mode considered but also on the sample-to-sample variability, represented by the $\xi^{(k)}$ variables. Though these fluctuations make the reconstruction of the average latent trajectory harder, they are actually helpful to recover the mode directions. 

The consequences of the measurement features that we have specifically introduced in our model are the following. The heterogeneity in the measurement noise strongly affects the quality in recovery of the components plagued by larger noise; sample-to-sample variability, is deleterious for the reconstruction of the signal trajectory, but beneficial for the inference of the modes; the distribution of rise and decay times of the kernel limits the recovery of the direction of the principal component, which can never be perfect and induces off-diagonal terms for $\rho$ even for very weak measurement noise.

Knowing how the errors on the reconstructed trajectories and modes depend both on the intrinsic noise in the dynamics and on the artifacts induced by the measurements may be important to decide what measurement methods should be preferred. One example is neuroscience, in which two main techniques to record neural population activities are currently employed, based on electrophysiology and on calcium imaging \cite{Brette2012}. These two approaches show different capabilities in terms of spatial and temporal resolutions, on the number of neurons they are able to record, and their own artifacts \cite{methods_neuro}. Recently, the activity in the motor cortex of behaving mice was recorded with these two approaches, leading to quite different inferred latent trajectories and modes \cite{Wei2020}. It is therefore crucial to better understand how measurement artifacts affect these results. We hope that the present work will be helpful in this regard. As mentioned above, the presence of component-dependent fluctuations in the convolution kernels has deep consequence for the quality of the inference with calcium-imaging methods. In particular, we observe that, even for vanishingly small measurement noise, the latent modes cannot be perfectly reconstructed.

Our study could be extended along several directions. First, in order to be applied to real data analysis, it should be completed with an inference procedure, capable of extracting the model parameters, such as $\sigma, \Delta, \Xi$, ... from a set of measurements. The explicit formula for the ground state energy $E_{GS}$ in terms of those parameters we were able to obtain makes this inference possible with gradient descent technique. We plan to report in a forthcoming publication an inference procedure, as well as some applications to real data, in particular in the context of neuroscience.  

Second, we have assumed that the measured quantities were linearly dependent upon the latent variables. While this hypothesis is at the core of PCA, it would be interesting to allow for non-linear effects in the data model, {\em e.g.} by applying a transfer function $\Phi$ onto $s$ in Eq.~\eqref{eq:dataraw}. The presence of non-linearities would make exact analytical calculations more challenging, but still feasible, as long as the number of modes, $K$, remains finite while $N,T\to\infty$ \cite{Reimann1996}. In this context, another interesting extension would be, in the present linear framework, to consider the case of extensive $K\propto N$. In such a setup, if the amplitudes of the latent coordinates $x^{(k)}$ decrease fast enough with the mode number $k$, we expect this problem to be analytically tractable. Such a calculation would permit one to understand how the effective number of recovered modes adapt to the quality and quantity of the available data. Last of all, let us mention that our approach could be extended to general linear auto-encoders, in which the encoder and decoder are not simply transposed of one another as is the case for PCA. A particularly interesting example is Demixed PCA, in which linear dimensional reduction is done in a semi-supervised way using annotations of external conditions \cite{Kobak2016}.

\vskip .3cm
\noindent
{\bf Acknowledgements.} M.L.'s PhD was funded by the 2021 CNRS 80 Prime initiative and by the ANR-21-CE16-0037 Locomat  project. L.B. and R.M. acknowledge fundings from the ANR-19-CE37-0016 Alert project.

\bibliographystyle{ieeetr}
\bibliography{main.bib}

\begin{thebibliography}{10}

\bibitem{Joliffe2002}
I.~Joliffe, {\em Principal Component Analysis}.
\newblock Springer-Verlag, 2002.

\bibitem{Jolliffe2016}
I.~T. Jolliffe and J.~Cadima, ``Principal component analysis: a review and
  recent developments,'' {\em Philosophical Transactions of the Royal Society
  A: Mathematical, Physical and Engineering Sciences}, vol.~374, no.~2065,
  p.~20150202, 2016.

\bibitem{Greenacre2022}
M.~Greenacre, P.~Groenen, and T.~e.~a. Hastie, ``Principal component
  analysis,'' {\em Nat Rev Methods Primers}, vol.~2, p.~100, 2022.

\bibitem{Rodarmel2002}
C.~Rodarmel and J.~Shan, ``component analysis for hyperspectral image
  classification,'' {\em Surv. Land. Inf. Sci}, vol.~62, pp.~115--122, 2002.

\bibitem{Du2007}
Q.~Du and J.~Fowler, ``Hyperspectral image compression using jpeg2000 and
  principal component analysis,'' {\em IEEE Geosci. Remote. Sens. Lett.},
  vol.~4, p.~201–205, 2007.

\bibitem{Paul2012}
L.~Paul and A.~Suman, ``Face recognition using principal component analysis
  method,'' {\em Int. J. Adv. Res. Comput. Eng. Technol.}, vol.~1,
  p.~135–139, 2012.

\bibitem{Ghorbani2020}
M.~Ghorbani and E.~K.~P. Chong, ``Stock price prediction using principal
  components,'' {\em PLoS One}, vol.~15, p.~e0230124, 2020.

\bibitem{Cunningham2014}
J.~P. Cunningham and B.~M. Yu, ``Dimensionality reduction for large-scale
  neural recordings,'' {\em Nature Neuroscience}, vol.~17, p.~1500–1509, Aug.
  2014.

\bibitem{Gallego2017}
J.~A. Gallego, M.~G. Perich, L.~E. Miller, and S.~A. Solla, ``Neural manifolds
  for the control of movement,'' {\em Neuron}, vol.~94, p.~978–984, June
  2017.

\bibitem{Williamson2019}
R.~Williamson, B.~Doiron, M.~Smith, and B.~Yu, ``Bridging large-scale neuronal
  recordings and large-scale network models using dimensionality reduction,''
  {\em Curr. Opin. Neurobiol.}, vol.~55, pp.~40--47, 2019.

\bibitem{Abraham2014}
G.~Abraham and M.~Inouye, ``Fast principal component analysis of large-scale
  genome-wide data,'' {\em PLoS One}, vol.~9, p.~e93766, 2014.

\bibitem{Alter2000}
O.~Alter, P.~O. Brown, and D.~Botstein, ``Singular value decomposition for
  genome-wide expression data processing and modeling,'' {\em Proc. Natl Acad.
  Sci.}, vol.~97, p.~10101–10106, 2000.

\bibitem{Tsuyuzaki2020}
K.~Tsuyuzaki, H.~Sato, K.~Sato, and I.~Nikaido, ``Benchmarking principal
  component analysis for large-scale single-cell rna-sequencing,'' {\em Genome
  Biol.}, vol.~21, p.~9, 2020.

\bibitem{Obukhov1947}
A.~Obhukov, ``Statistically homogeneous fields on a sphere,'' {\em Usp. Mat.
  Navk.}, vol.~2, pp.~169--198, 1947.

\bibitem{Lorenz1956}
E.~Lorenz, ``Empirical orthogonal functions and statistical weather
  prediction,'' tech. rep., Statistical Forecast Project Report 1, Dept. of
  Meteor. MIT: 49, 1956.

\bibitem{Preisendorfer1988}
R.~Preisendorfer and C.~Mobley, {\em Principal component analysis in
  meteorology and oceanography}.
\newblock Amsterdam, The Netherlands: Elsevier, 1988.

\bibitem{Benzecri1973}
J.-P. Benzécri, {\em Analyse Des Données, Tome 2: Analyse Des
  Correspondances}.
\newblock Dunod, 1973.

\bibitem{Watkin1994}
T.~L.~H. Watkin and J.~P. Nadal, ``Optimal unsupervised learning,'' {\em
  Journal of Physics A: Mathematical and General}, vol.~27, p.~1899–1915,
  Mar. 1994.

\bibitem{Reimann1996}
P.~Reimann and C.~Van~den Broeck, ``Learning by examples from a nonuniform
  distribution,'' {\em Physical Review E}, vol.~53, p.~3989–3998, Apr. 1996.

\bibitem{hoyle2004}
D.~C. Hoyle and M.~Rattray, ``Principal-component-analysis eigenvalue spectra
  from data with symmetry-breaking structure,'' {\em Phys. Rev. E}, vol.~69,
  p.~026124, Feb 2004.

\bibitem{Johnstone2001}
I.~M. Johnstone, ``On the distribution of the largest eigenvalue in principal
  components analysis,'' {\em Ann. Stat.}, vol.~29, pp.~295--327, Apr. 2001.

\bibitem{Baik2004}
J.~{Baik}, G.~B. {Arous}, and S.~{Peche}, ``Phase transition of the largest
  eigenvalue for non-null complex sample covariance matrices,'' {\em The Annals
  of Probability}, 2004.

\bibitem{Oba2007}
S.~Oba, M.~Kawanabe, K.-R. M\"{u}ller, and S.~Ishii, ``Heterogeneous component
  analysis,'' in {\em Advances in Neural Information Processing Systems}
  (J.~Platt, D.~Koller, Y.~Singer, and S.~Roweis, eds.), vol.~20, Curran
  Associates, Inc., 2007.

\bibitem{Shi2024}
N.~Shi and R.~A. Kontar, ``Personalized pca: Decoupling shared and unique
  features,'' {\em Journal of Machine Learning Research}, vol.~25, no.~41,
  pp.~1--82, 2024.

\bibitem{Steinberg2024}
J.~Steinberg, U.~Adomaitytė, A.~Fachechi, P.~Mergny, D.~Barbier, and
  R.~Monasson, ``Replica method for computational problems with randomness:
  principles and illustrations,'' {\em Journal of Statistical Mechanics: Theory
  and Experiment}, p.~104002, oct 2024.

\bibitem{Legenkaia2024}
M.~Legenkaia, {\em Reconstruction of low-dimensional neural trajectories from
  population activity recordings: from statistical limitations to experimental
  design}.
\newblock PhD thesis, Ecole Normale Sup\'erieure-PSL, 2024.

\bibitem{newton}
J.~Nocedal and S.~Wright, {\em Numerical Optimization}.
\newblock Springer Series in Operations Research and Financial Engineering
  (Springer New York) ISBN 9780387303031, 2006.

\bibitem{Note1}
The same behaviour can be observed for $\epsilon $, see Figure~\ref
  {fig:epsilon2}.

\bibitem{Brette2012}
R.~Brette and A.~Destexhe, {\em Intracellular recording}, p.~44–91.
\newblock Cambridge University Press, Sept. 2012.

\bibitem{methods_neuro}
K.~Harris, R.~Quiroga, J.~Freeman, and S.~Smith, ``Improving data quality in
  neuronal population recordings,'' {\em Nat. Neurosci.}, vol.~19,
  pp.~1165--74, 2016.

\bibitem{Wei2020}
Z.~Wei, B.-J. Lin, T.-W. Chen, K.~Daie, K.~Svoboda, and S.~Druckmann, ``A
  comparison of neuronal population dynamics measured with calcium imaging and
  electrophysiology,'' {\em PLOS Computational Biology}, vol.~16, p.~e1008198,
  Sept. 2020.

\bibitem{Kobak2016}
D.~Kobak, W.~Brendel, C.~Constantinidis, C.~E. Feierstein, A.~Kepecs, Z.~F.
  Mainen, X.-L. Qi, R.~Romo, N.~Uchida, and C.~K. Machens, ``Demixed principal
  component analysis of neural population data,'' {\em eLife}, vol.~5,
  p.~e10989, apr 2016.

\end{thebibliography}

\end{document}